\documentclass[a4paper,10pt]{article}
\usepackage[utf8x]{inputenc}
\usepackage{graphicx}
\usepackage{setspace}

\title{Astronomical Dating and the Internal Chronology of the Pentateuch}
\author{Pierfrancesco La Mura\\
\texttt{plamura@hhl.de}}

\begin{document}

\maketitle

\begin{abstract}
Some of the narratives in the Pentateuch can be associated with known astronomical events to provide absolute dates for biblical chronology.
\end{abstract}

\section{Introduction}

Early, confessional attempts aimed at establishing a consistent biblical chronology, such as the Seder Olam Rabbah, strove to identify the intended timeline of events narrated in the Pentateuch purely from information contained in the text, such as genealogical data and narrative clues. While internally consistent, the resulting chronologies were ultimately incompatible with a growing body of scientific and historical knowledge: for instance, the seventeenth-century bishop James Ussher calculated that God created the universe on October 22, 4004 BCE [1], an untenable thesis in the light of modern physical cosmology. 

The Pentateuch's own insistence on timekeeping suggests that a precise timeline for at least some of the events was available at the time the text was written. Yet, dauntingly, the genealogical and biographical information contained in it proved insufficient to identify a unique, reliable chronology, and as a consequence few, if any, of the narrated events could be provided with a firm connection to independent historical or archaeological evidence. 

Modern, non-confessional approaches have variously attempted to correlate the narratives with external information, such as historical records or archaeological evidence, in order to identify a reliable timeline for the events described in the Pentateuch. Yet, the introduction of external elements has often been carried out at the expense of consistency with internal textual constraints: for instance, the birth of agriculture in the Near East according to archaeological evidence seems irreconcilable with  the dating implied by the genealogies of Genesis.

\bigskip

As we shall see the Pentateuch does contain a complete and reliable internal timekeeping system, but its reckoning of time is not limited to time spans: it also involves astronomical information, reflecting the high status of the study of astronomy in Judaism. 

\begin{quote}
 R. Simeon b. Pazzi said in the name of R. Joshua b. Levi on the authority of Bar Kappara: He who knows how to calculate the cycles and planetary courses, but does not, of him Scripture saith, but they regard not the work of the Lord, neither have they considered the operation of his hands [Isa 5:12]. R. Samuel b. Nahmani said in R. Johanan's name: How do we know that it is one's duty to calculate the cycles and planetary courses? Because it is written, for this is your wisdom and understanding in the sight of the peoples [Deu 4:6]: what wisdom and understanding is in the sight of the peoples? Say, that it is the science of cycles and planets.
\begin{flushright}
Babylonian Talmud, Shab. 75a\end{flushright}
\end{quote} 

The origins of Western astronomy can be traced to Mesopotamia [2]. In the earliest Babylonian star catalogues, dating from about 1200 BCE, many star names appear in Sumerian, suggesting a continuity reaching into the Early Bronze Age. The above passage from the Babylonian Talmud of course represents a much later tradition: its interest in our context comes from the fact that it emphasizes the duty to calculate, as opposed to simply observe or record, astronomical events. Another element which also points to a calculational perspective is the observation that, in some manuscripts of the Samaritan Targum, the word ``astrolabe'' appears in place of ``idols'' in the context of Jacob's escape from Laban's house. In our analysis of Pentateuch chronology a calculational - as opposed to observational - perspective, as could have been afforded at the time of redaction by the use of an astrolabe, will turn out to be essential.

\bigskip

As we shall strive to demonstrate, nearly all of the astronomical descriptions which are found in the text refer to solar eclipses and close passages of the Sun and the Moon at the Northern (also known as Spring, Vernal, or Paschal) equinox. In the Northern hemisphere the Vernal equinox is closely associated with the renewal of life, and in several calendar systems of the Near East marks the beginning of the new year. In particular, whenever it falls on a new moon, the Vernal equinox also marks the beginning of the new year in the Jewish religious calendar (Nisan 1)\footnote{Solar eclipses only occur with a new moon, and hence in the Jewish calendar they may only take place at the beginning of each lunar month.}.

In those descriptions, we find that the Sun typically represents the ``glory of God'': directly, or symbolically through the current protagonist of the narrative or his firstborn. The new Moon, which in a solar eclipse appears to partially or totally cover the Sun, variously represents the ``hand of God'', the hand of the protagonist, or that of a third party; but also a rainment, the mouth of a well, or the Egyptian bondage. The ecliptic, along which the constellations of the Zodiac are found, is a caravan circle, a bow, a sword, or a ladder. 
Eclipses which, from the point of view of an observer located in the Near East, take place at sunrise or sunset are often used to identify a geographical feature at the visible horizon, for instance a town or mountain.

Approximate alignments of the Sun and the Moon on the same calendar day follow the 19-year Metonic cycle. The cycle was presumably well known to the author of the Pentateuch, as the nineteenth verse in Genesis 1 is the conclusion of the fourth day, in which the Sun and the Moon are created. Also worth of mention, in the light of our proposed identification of the Sun and the Moon as the glory and the hand of God, respectively, is the opening of Psalm 19: ``The heavens declare the glory of God; the skies proclaim the work of his hands''.\footnote{On the astronomical interpretation of Psalm 19 see A. Lenzi, "The Metonic Cycle, Number Symbolism, and the Placement of Psalms 19 and 119 in the MT Psalter", JSOT 24.4, 2010.} 

\bigskip

The analysis will proceed as follows. Taking Noah's Flood as the starting point, and proceeding until the death of Moses, we shall identify a sequence of passages in the text, and demonstrate that each passage can be associated to a specific solar eclipse, in such a way that: (i) all solar eclipses at the Northern equinox in the reconstructed timelines of the narratives are accounted for; (ii) all occurrences of the words ``appeared'' and ``covenant'' in the corresponding portions of the text are accounted for. 

Next, we shall turn to the genealogies of the pre-flood patriarchs: for each of the three main extant manuscript families (Masoretic, Septuagint and Samaritan), we test the hypotesis that the corresponding chronologies are independent of eclipse events. As we shall see, using a simple binomial model, while for the Masoretic and Septuagint chronologies independence cannot be rejected at any interesting significance level, the hypothesis that the dates in the reconstructed Samaritan chronology were set independently of eclipse events is rejected at the 0.04\% significance level.

\bigskip

For convenience, we shall sometimes express dates in astronomical notation. In such notation the year -1000 corresponds to 1001 BCE, reflecting the fact that there is no year 0 in the Gregorian calendar; furthermore, all dates before October 15, 1582 are conventionally expressed as Julian. We use Stellarium 0.10.6 (available as a free download at www.stellarium.org) to reconstruct the position of the celestial bodies at specific times and places. 

\bigskip

Finally, we shall refer to all eclipses which admit a totality (respectively, annularity) path as total (respectively, annular); we refer to all other eclipses as partial. For simplicity, in case an eclipse is both annular and total we record it as annular. Otherwise, whenever applicable, our classification coincides with the one in NASA's Five Millennium Catalog of Solar Eclipses\footnote{NASA's Five Millennium Catalog of Solar Eclipses (-1999 to 3000),

http://eclipse.gsfc.nasa.gov/SEcat5/catalog.html.}. 
Furthermore, we shall refer to those events in which, from the point of view of a suitably located Earth observer, the Sun and the Moon appear within one solar disk from each other as close passages. For each event we also report whether, from the point of view of a putative observer located in the Near East, it would have occurred during daytime or nighttime. In almost all cases, we conventionally take Jerusalem as the reference point. 

\section{The Great Flood}

William Ryan and Walter Pitman suggested in 1999 that several related Near Eastern Flood stories, including Noah's, could be associated with the sudden and perhaps violent flooding of the Black Sea region in the middle of the 6th millennium BCE [3,4]. The analysis of geologic and organic sediments on the Black Sea floor revealed ancient shorelines and deltas, and the abrupt disappearance around 5500 BCE of freshwater mollusks, replaced by marine species. According to Ryan and Pitman, around that time the waters of the Mediterranean, whose level had been increasing by over 50 meters since the beginning of the Holocene, spilled over the Bosphorus and flooded the vast plains and shorelines surrounding what at the time was a large fresh- or brackish- water lake. 

\bigskip

\begin{verse}
Gen 6:12 And God looked upon the earth, and, behold, it was corrupt [\textit{lit.}: smeared]; for all flesh had corrupted [\textit{lit.}: smeared] his way upon the earth.
\end{verse}

\begin{verse}
Gen 6:18 But with thee will I establish my covenant; and thou shalt come into the ark, thou, and thy sons, and thy wife, and thy sons' wives with thee.
\end{verse}

\begin{quote}
-5504/5/4 

Nighttime, partial eclipse at the Northern equinox. The Sun in Gemini (the brotherhood of mankind, as the ``glory of God''), following its path upon the surface of the Earth, smears it with the shadow of the Moon (by its own hand). 
\end{quote}

The event, which we provisionally associate with the equinoxial eclipse of -5504, takes place just before the Flood, as Noah begins to build the Ark. It announces, via the 19-year Metonic cycle, another nighttime, total eclipse which takes place on -5485/5/5, 
 shortly after the end of the Flood. 
\bigskip

\begin{verse}
Gen 9:13-14 I do set my bow in the cloud, and it shall be for a token of a covenant between me and the earth. And it shall come to pass, when I bring a cloud over the earth, that the bow shall be seen in the cloud.
\end{verse} 

The bow in the cloud not only symbolizes the rainbow, but also the ecliptic: in which case, whenever the Moon (as the ``hand of God'') with its shadow brings a cloud over the Earth, it also appears to touch the ecliptic (the bow). Knowledge of the ecliptic allows for a lunisolar calendar, which is kept in tune by direct astronomical observation of the equinoxial points. The additional knowledge of the Metonic cycle also affords some ability to predict the occurrence of eclipses, which in turn, as the passage suggests, come to be regarded as a symbol of covenant. 

\bigskip

\begin{verse}
Gen 9:23 And Shem and Japheth took a garment, and laid [it] upon both their shoulders, and went backward, and covered the nakedness of their father; and their faces [were] backward, and they saw not their father's nakedness.\end{verse} 

\begin{center}
 \includegraphics[width=340pt,keepaspectratio=true]{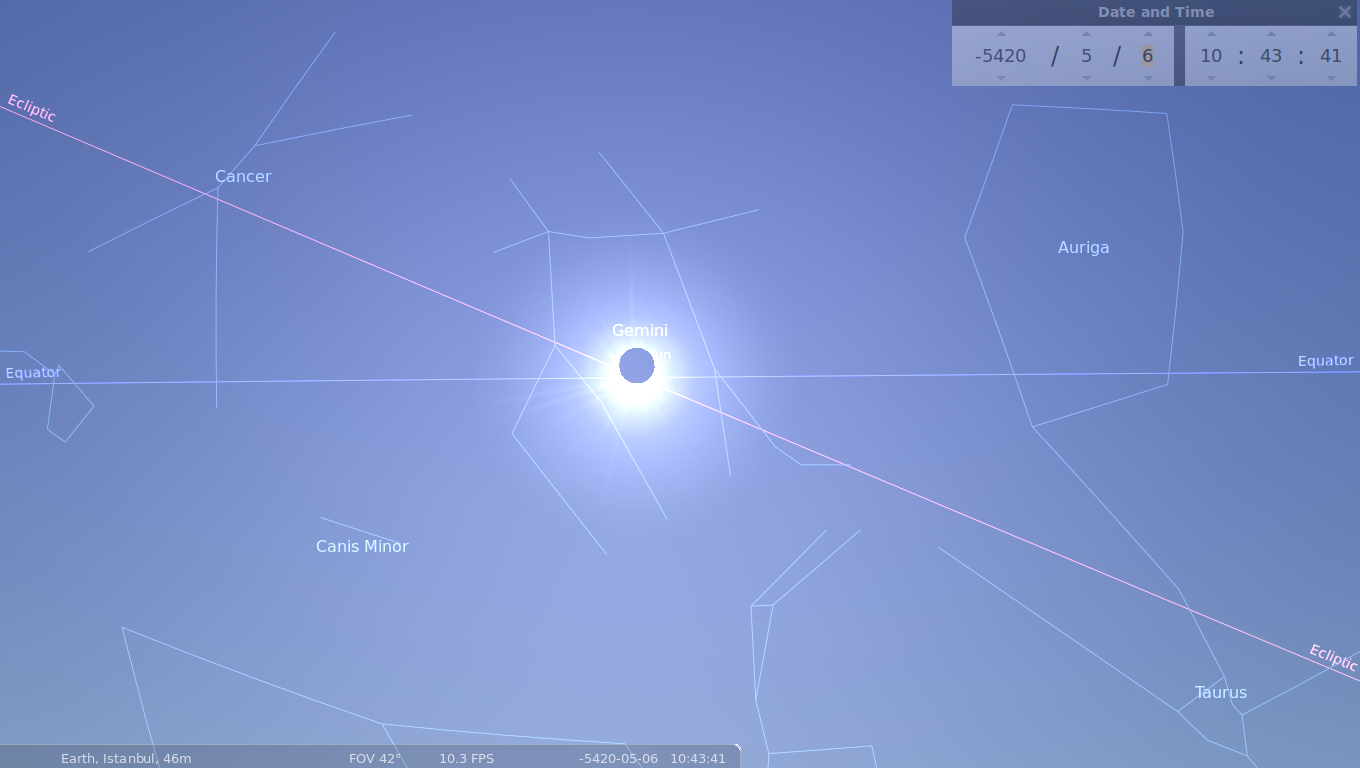}
\end{center}

\begin{quote}
-5420/5/6 

Daytime, partial eclipse at the Northern equinox. The Sun in Gemini is covered by the Moon. The Twins appear to walk backward, escorting the Sun along the ecliptic. \end{quote} 

The equinoxial points slowly rotate with respect to the constellations of the Zodiac, returning to the same position approximately every 25,800 years. As a consequence, no other fully daytime (in the Near East) solar eclipse at the Northern equinox in the last 30,000 years had the Sun standing just in the middle of Gemini. 

\bigskip

Note that, here and in other descriptions of astronomical events, the position of the relevant constellations was of course not directly observable as it was daytime, and could only be deduced through astronomical calculations. 

\section{Chronology of Abraham}

The genealogies of the post-flood patriarchs before Abram, like those of their pre-flood predecessors, are comprised of two figures per patriarch. The table below summarizes the relevant figures, according to the Masoretic text.

\bigskip

\begin{center}
\begin{tabular}{lll}
Shem & 100 & 500\\
Arphaxad & 35 & 403\\
Salah & 30 & 403\\
Eber & 34 & 430\\
Peleg & 30 & 209\\
Reu & 32 & 207\\
Serug & 30 & 200\\
Nahor & 29 & 119\\
Terah & 70 & 205\\
\end{tabular}
\end{center}

\bigskip

The opening of Genesis 5 suggests that the name of each patriarch, starting with Adam, is representative of a house, or lineage.

\begin{verse}
Gen. 5:2-3 Male and female created he them; and blessed them, and called their name Adam, in the day when they were created. And Adam lived an hundred and thirty years, and begat a son in his own likeness, and after his image; and called his name Seth.\end{verse}

We regard the name of each patriarch as representative of a house or dynasty, and the total years of his life as reignal years. For instance, Eber lived a total of 34 + 430 years. Hence, Peleg is born in the 34th year of Eber, but the years of his life (reign) only begin to be counted 430 years later, when he succeeds Eber\footnote{Observe that, in the counting conventions of the Pentateuch, the first year of life (reign) is already counted as year one: see  Gen 7:6 ``And Noah [was] six hundred years old when the flood of waters was upon the earth.'', and Gen 7:11 ``In the six hundredth year of Noah's life, in the second month, the seventeenth day of the month, the same day were all the fountains of the great deep broken up, and the windows of heaven were opened.'' Hence, when Eber was 34 years old, he was also in his 34th year.}.

\bigskip

Adding up the dynastic ages of the patriarchs from the time of the Flood to the beginning of Abram's life, and considering that Arphaxad was born 2 years after the Flood (Gen 11:10), gives 2676 + 290 + 2 = 2968 years.


\bigskip

Considering that -5420 was long enough after the Flood that Noah had become a husbandman, planted a vineyard and started to make wine (Gen 9:20), let us conventionally set that timespan to 80 years, and hence the Flood at -5500\footnote{It is worth noting that, according to the early-medieval Book of the Rolls, ''When Reu was thirty-two years old, Serug was born to him; the length of his life being 232 years. At the end of 163 years of the life of Reu, Nimrod the giant reigned over the whole earth. The beginning of his kingdom was from Babel. It was he who saw in the sky a piece of black cloth and a crown". If the Flood took place in -5500, then the beginning of Nimrod's reign was in -3261. In the same year there was an annular eclipse at the Northern equinox (a "black cloth, and a crown").}. As it turns out, -5500 is the only date which is fully consistent with the chronology of later events. 

\bigskip

If the Flood took place in -5500, then Abram began his life (reign) 2968 years later, in -2532. Abram is 75 years old in -2457, when he departs from Haran (Gen 12:4).


\bigskip

\begin{verse}
Gen 12:1 Now the LORD had said unto Abram, Get thee out of thy country, and from thy kindred, and from thy father's house, unto a land that I will shew thee.\end{verse} 

\begin{verse}
Gen 12:6-8 And Abram passed through the land unto the place of Sichem, unto the plain of Moreh. And the Canaanite [was] then in the land. And the LORD appeared unto Abram, and said, Unto thy seed will I give this land: and there builded he an altar unto the LORD, who appeared unto him. And he removed from thence unto a mountain on the east of Bethel, and pitched his tent, [having] Bethel on the west, and Hai on the east: and there he builded an altar unto the LORD, and called upon the name of the LORD.\end{verse} 

\begin{center}
 \includegraphics[width=340pt,keepaspectratio=true]{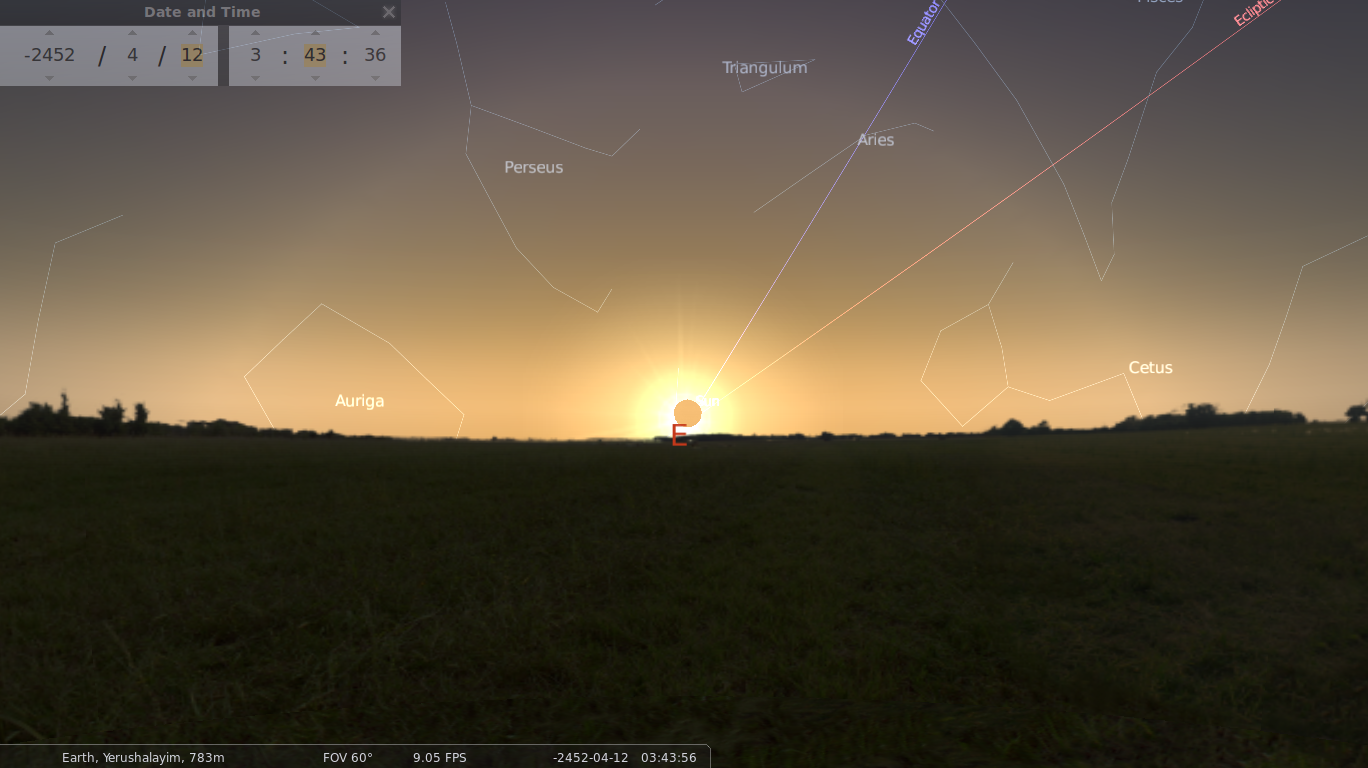}
\end{center}

\begin{quote}
-2452/4/12

Northern equinox, sunrise. A total eclipse of the Sun marks the eastern horizon, predicted via the 19-year Metonic cycle by an analogous event which had taken place on -2471/4/12. 
\end{quote} 

Abram builds a first altar in the place where he witnessed the omen, and then builds a second altar, presumably at the eastern horizon of the first, to mark the land's borders. In -2452 Abram is exactly 80 years old.

\bigskip

\begin{verse}
Gen 15:9-10 And he said unto him, Take me an heifer of three years old, and a she goat of three years old, and a ram of three years old, and a turtledove, and a young pigeon. And he took unto him all these, and divided them in the midst, and laid each piece one against another: but the birds divided he not.
\end{verse}

\begin{verse}
Gen 15:17-18 And it came to pass, that, when the sun went down, and it was dark, behold a smoking furnace, and a burning lamp that passed between those pieces. In the same day the LORD made a covenant with Abram, saying, Unto thy seed have I given this land, from the river of Egypt unto the great river, the river Euphrates.
\end{verse} 

\begin{quote}
-2441/4/11

Northern equinox, close passage at nighttime. The Sun (a burning lamp) and the new Moon (a smoking furnace) come close to each other while passing between Aries (ram and goat), Taurus (a heifer) and the Pleiades (pigeon and dove). 
\end{quote}

At the time of the omen Abram was 91 years old.

\bigskip

\begin{verse}
Gen 17:10 This [is] my covenant, which ye shall keep, between me and you and thy seed after thee; Every man child among you shall be circumcised.\end{verse} 

\begin{verse}
Gen 18:1-2 And the LORD appeared unto him in the plains of Mamre: and he sat in the tent door in the heat of the day; And he lift up his eyes and looked, and, lo, three men stood by him.\end{verse} 

\begin{center}
 \includegraphics[width=340pt,keepaspectratio=true]{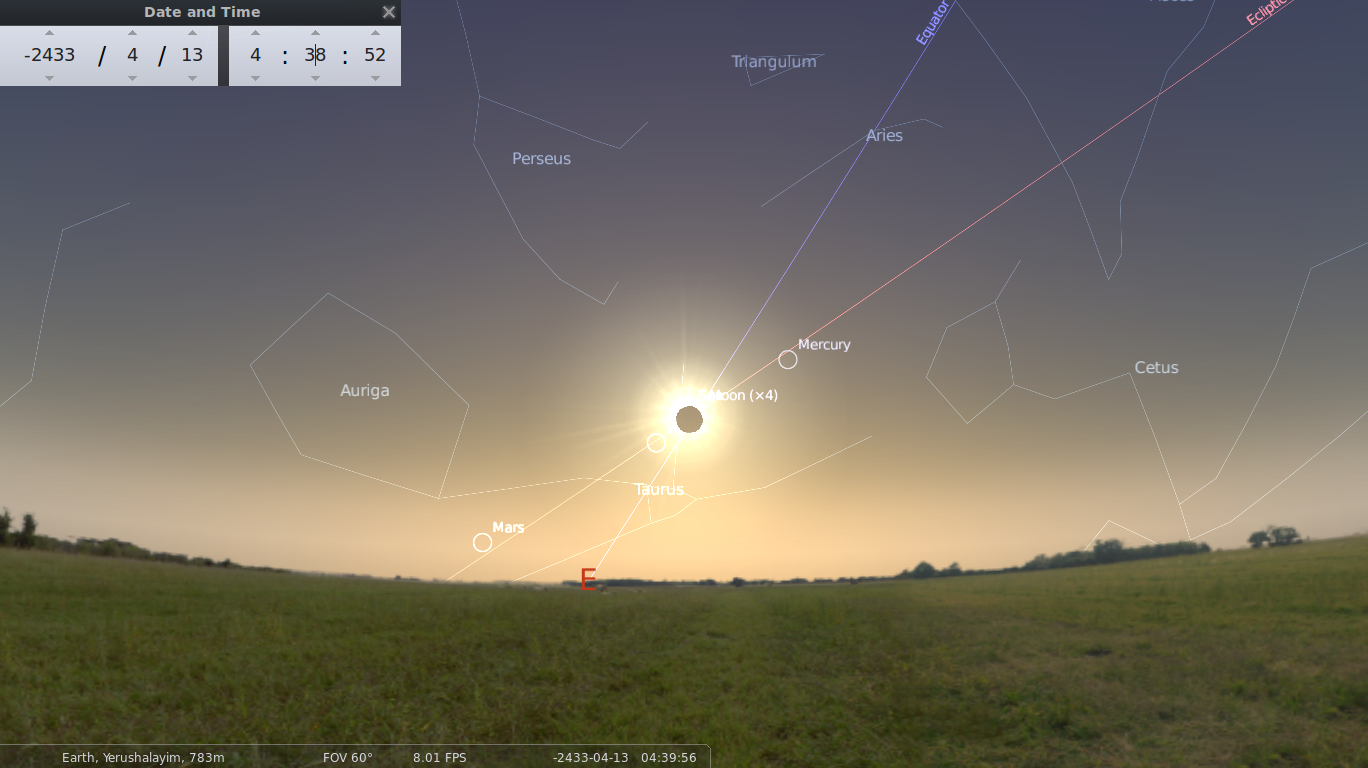}
\end{center}

\begin{quote}
-2433/4/13

Northern equinox, annular eclipse at sunrise. The Moon (the ``hand of God'') cuts out a ring from the Sun's disk (the ``glory of God''). Three planets stand by: Mars, Venus and Mercury.\end{quote} 

In -2433 Abraham is 99 years old. One year later, in -2432, Isaac is born (Gen 21:5).

\bigskip

\begin{verse}
Gen 21:25-28 And Abraham reproved Abimelech because of a well of water, which Abimelech's servants had violently taken away. And Abimelech said, I wot not who hath done this thing: neither didst thou tell me, neither yet heard I [of it], but to day. And Abraham took sheep and oxen, and gave them unto Abimelech; and both of them made a covenant. And Abraham set seven ewe lambs of the flock by themselves.
\end{verse} 

\begin{center}
 \includegraphics[width=340pt,keepaspectratio=true]{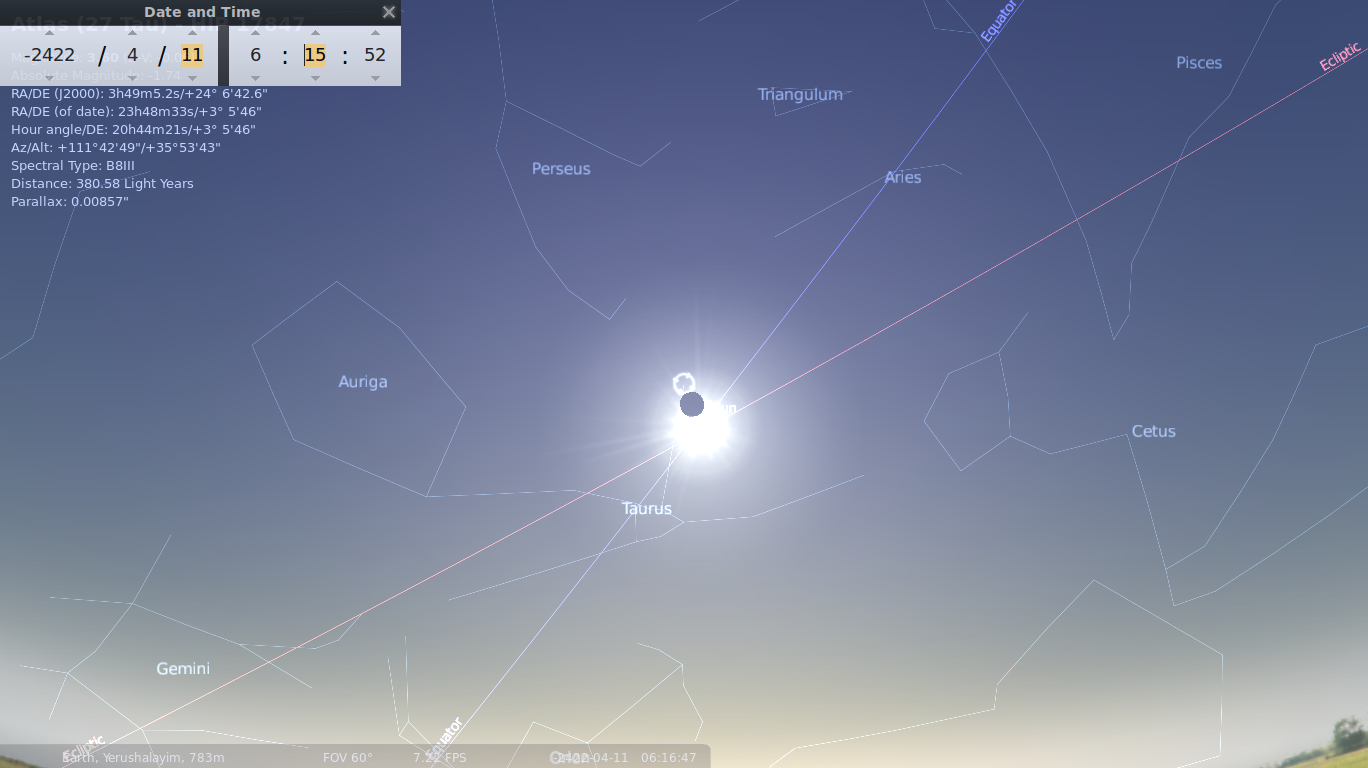}
\end{center}

\begin{quote}
-2422/4/11

At the Northern equinox, the Sun (Abraham) moves along the ecliptic, carrying with it Aries and Taurus (``sheep and oxen''), while the Moon (his hand) singles out the Pleiades (``seven  ewe lambs''). 
 \end{quote} 

In -2422 Abraham is exactly 110 years old.

\bigskip

\begin{verse}
Gen 22:11-13 And the angel of the LORD called unto him out of heaven, and said, Abraham, Abraham: and he said, Here [am] I. And he said, Lay not thine hand upon the lad, neither do thou any thing unto him: for now I know that thou fearest God, seeing thou hast not withheld thy son, thine only [son] from me. And Abraham lifted up his eyes, and looked, and behold behind [him] a ram caught in a thicket by his horns.
\end{verse} 

\begin{center}
 \includegraphics[width=340pt,keepaspectratio=true]{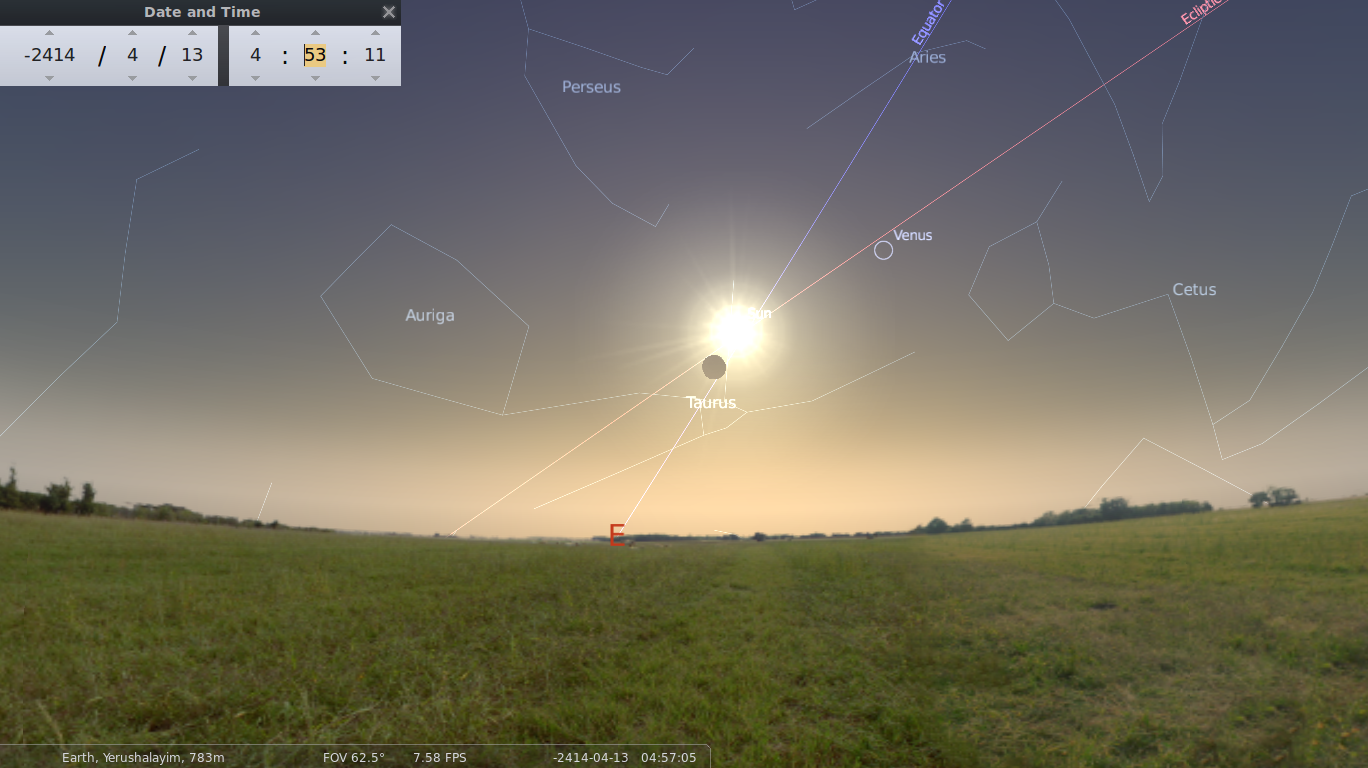}
\end{center}

\begin{quote}
-2414/4/13

Northern equinox, annular eclipse just before sunrise, predicted via the 19-year Metonic cycle by the -2433 eclipse announcing Isaac's birth. At sunrise, the Sun in Taurus (Isaac, as the ``glory of God'', in the role of sacrificial ox) is bound at the crossing of the ecliptic and the celestial equator, while the Moon (Abraham, as the ``hand of God'') slowly backs away from it. In the background, Aries (``a ram'') touches the celestial equator with its horn (the star Hamal). Abraham is 118 years old at the time of the omen. 
\end{quote} 

The sixth-century synagogue of Beth Alpha includes among its mosaics a zodiac wheel, and a depiction of the binding of Isaac [15]. Just below the line of the horizon, while Abraham approaches the bound Isaac, a radiating disc is covered by a dark silhouette, from which protrudes the hand of God, carrying the inscription ''do not raise".

\bigskip

\begin{verse}
Gen 22:15 And the angel of the LORD called unto Abraham out of heaven the second time.\end{verse}

\begin{quote}
-2406 4/14 

Northern equinox, close passage at sunset. Now Abraham is 126 years old. \end{quote} 

\bigskip

\begin{verse}
Gen 23:1-2 And Sarah was an hundred and seven and twenty years old: [these were] the years of the life of Sarah. And Sarah died in Kirjatharba; the same [is] Hebron in the land of Canaan.\end{verse} 

\begin{verse}
Gen 23:3 And Abraham stood up from before his dead, and spake unto the sons of Heth.\end{verse} 

\begin{quote}
-2395 4/12

Northern equinox. Partial eclipse around noon, predicted by the Metonic cycle. In -2395 Abraham is exactly 137 years old (see Gen 17:17, Gen 23:1). \end{quote} 

\begin{verse}
Gen 23:7 And Abraham stood up, and bowed himself to the people of the land, [even] to the children of Heth.\end{verse} 

\begin{quote}
-2387 4/14

Northern equinox. Annular eclipse at sunrise, predicted by the Metonic cycle. Unveiling of Sarah's tombstone, 8 years after her death.\end{quote} 

\begin{verse}
Gen 23:12 And Abraham bowed down himself before the people of the land.\end{verse} 

\begin{quote}
-2368 4/13

Northern equinox. Annular eclipse at sunset, predicted by the Metonic cycle, and also the last equinoxial event in Abraham's timeline. Dedication of the cave in Machpelah, 19 years after the death of Sarah.\end{quote}

\bigskip

Appendix 1 contains a list of all solar eclipses at the Northern equinox in the period -2862 to -2061. The sequence of astronomical events witnessed by Abraham, constrained by the genealogical and biographical information provided in the text, is uniquely identified within that time span.

\section{The Fear of Isaac}

If Abram began his life (reign) in -2532 then Isaac was born in -2432, and began his reign 75 years later, in -2357 (Gen 25:7).

\begin{verse}
Gen 21:4 And Abraham circumcised his son Isaac being eight days [\textit{lit. ``yowm``}: days, years] old, as God had commanded him.\end{verse} 

\begin{quote}
-2349 4/13 

Northern equinox. Annular eclipse at nighttime, predicted by the Metonic cycle. Isaac is now 8 years old. Isaac is 40 years old in -2317, when he marries Rebekah (Gen 25:20).\end{quote} 

\bigskip

\begin{verse}
Gen 26:1-2 And there was a famine in the land, beside the first famine that was in the days of Abraham. And Isaac went unto Abimelech king of the Philistines unto Gerar. And the LORD appeared unto him, and said, Go not down into Egypt; dwell in the land which I shall tell thee of.\end{verse} 

\begin{quote}
-2303/4/14

Northern equinox. Total eclipse at nighttime, predicted by the Metonic cycle. Now Isaac is 54 years old. Jacob is born 6 years later, in -2297 (Gen 25:26).\end{quote} 

\bigskip

At the time of Isaac's famine, Sargon of Akkad held a vast empire which is known to have extended from Elam to the Mediterranean sea, including Mesopotamia, parts of modern-day Iran and Syria, and possibly parts of Anatolia and the Arabian peninsula. Sargon reigned from 2334 to 2279 BC (short chronology) [5]. The Chronicle of Early Kings reports that Sargon's empire suffered a famine during the latter years of his reign [6].

\bigskip

\begin{verse}
Gen 26:23-25 And he went up from thence to Beersheba. And the LORD appeared unto him the same night, and said, I [am] the God of Abraham thy father: fear not, for I [am] with thee, and will bless thee, and multiply thy seed for my servant Abraham's sake. And he builded an altar there, and called upon the name of the LORD, and pitched his tent there: and there Isaac's servants digged a well.                                                                                                                                                                                                                                                                                                                                                                                                         \end{verse} 

\begin{quote}
-2284/4/14

Northern equinox. Total eclipse at nighttime, predicted by the Metonic cycle. Now Isaac is 73 years old. Observe that there are in total four eclipses associated with Isaac, and all of them take place at night. \end{quote} 

\section{Birth of Israel}

\begin{verse}
Gen 28:12-13, And he dreamed, and behold a ladder set up on the earth, and the top of it reached to heaven: and behold the angels of God ascending and descending on it. And, behold, the LORD stood above it.\end{verse} 

\begin{verse}
Gen 28:16 And Jacob awaked out of his sleep, and he said, Surely the LORD is in this place; and I knew [it] not.\end{verse} 

\begin{center}
 \includegraphics[width=340pt,keepaspectratio=true]{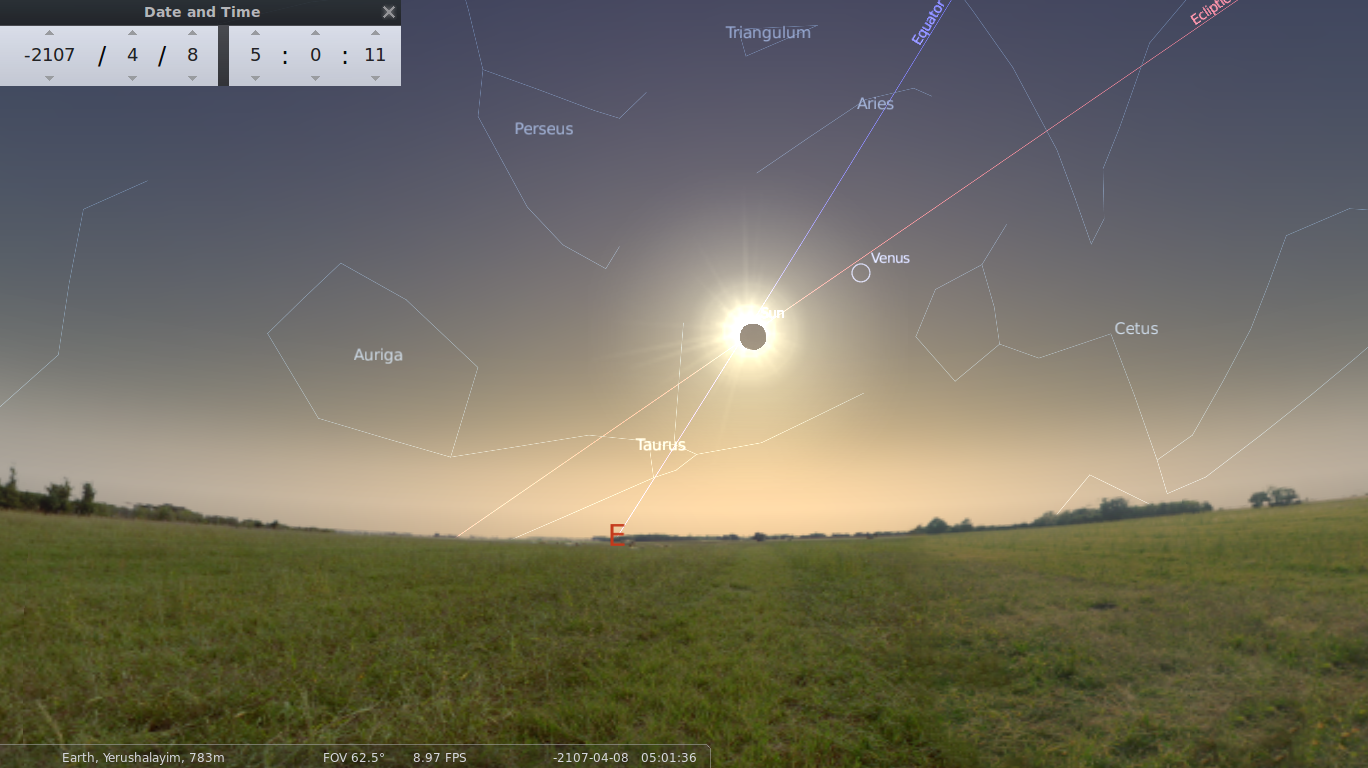}
\end{center}

\begin{quote}
-2107/4/8

Northern equinox, total eclipse at sunrise. The Sun (the glory of God) rises while covered by the Moon, standing on the ecliptic (the ladder), along with the constellations of the Zodiac (the angels of God). The Sun only appears in its full strength about two hours later than normal. \end{quote} 

If Abram began his life (reign) in -2532, interpreting the age of Abraham (175 years) and that of Isaac (180 years) as dynastic time spans, as we did with the other patriarchs, we can place the beginning of Jacob's life (reign) at -2177. Hence, in -2107 Jacob was 70 years old, and he was 84 years old in -2093 when Joseph was born.

\begin{center}
\begin{tabular}{ll}
Abram reigns & -2532\\
Isaac born & -2432\\
Isaac reigns & -2357\\
Jacob born & -2297\\
Jacob reigns & -2177
\end{tabular}
\end{center}

\begin{verse}
Gen 31:11-13 And the angel of God spake unto me in a dream, [saying], Jacob: And I said, Here [am] I. And he said, Lift up now thine eyes, and see, all the rams which leap upon the cattle [are] ringstraked, speckled, and grisled: for I have seen all that Laban doeth unto thee. I [am] the God of Bethel, where thou anointedst the pillar, [and] where thou vowedst a vow unto me: now arise, get thee out from this land, and return unto the land of thy kindred.\end{verse} 

\begin{quote}
-2099/4/9

Northern equinox. Close passage at sunset. Aries (the rams) preceeds the Sun, and Taurus (the cattle) follows it along the ecliptic.\end{quote} 

Jacob is 78 years old when he witnesses the omen. Observe that the name Laban (\textit{lit.}: white) is also the root of the Hebrew word for Moon, \textit{lbanah}. 
Moreover, observe that the terms ``ringstraked, speckled and grisled'' accurately describe all the possible ways in which the solar disc may appear ``stained'' by the shadow of the Moon during the different stages of an eclipse. They also anticipate the three subsequent equinoxial events, namely: a total eclipse, a partial one, and a close passage.

\bigskip

\begin{verse}
Gen 31:21-23 So he fled with all that he had; and he rose up, and passed over the river, and set his face [toward] the mount Gilead. And it was told Laban on the third day [\textit{lit.}: year] that Jacob was fled. And he took his brethren with him, and pursued after him seven days' [\textit{lit.}: years'] journey; and they overtook him in the mount Gilead. \end{verse} 

\begin{verse}
Gen 31:25-26 Then Laban overtook Jacob. Now Jacob had pitched his tent in the mount: and Laban with his brethren pitched in the mount of Gilead. And Laban said to Jacob, What hast thou done, that thou hast stolen away unawares to me, and carried away my daughters, as captives [taken] with the sword?
\end{verse} 

\begin{verse}
Gen 31:29 It is in the power of my hand to do you hurt: but the God of your father spake unto me yesternight, saying, Take thou heed that thou speak not to Jacob either good or bad.
\end{verse} 

\begin{verse}
Gen 31:43 And Laban answered and said unto Jacob, [These] daughters [are] my daughters, and [these] children [are] my children, and [these] cattle [are] my cattle.
\end{verse} 

\begin{center}
 \includegraphics[width=340pt,keepaspectratio=true]{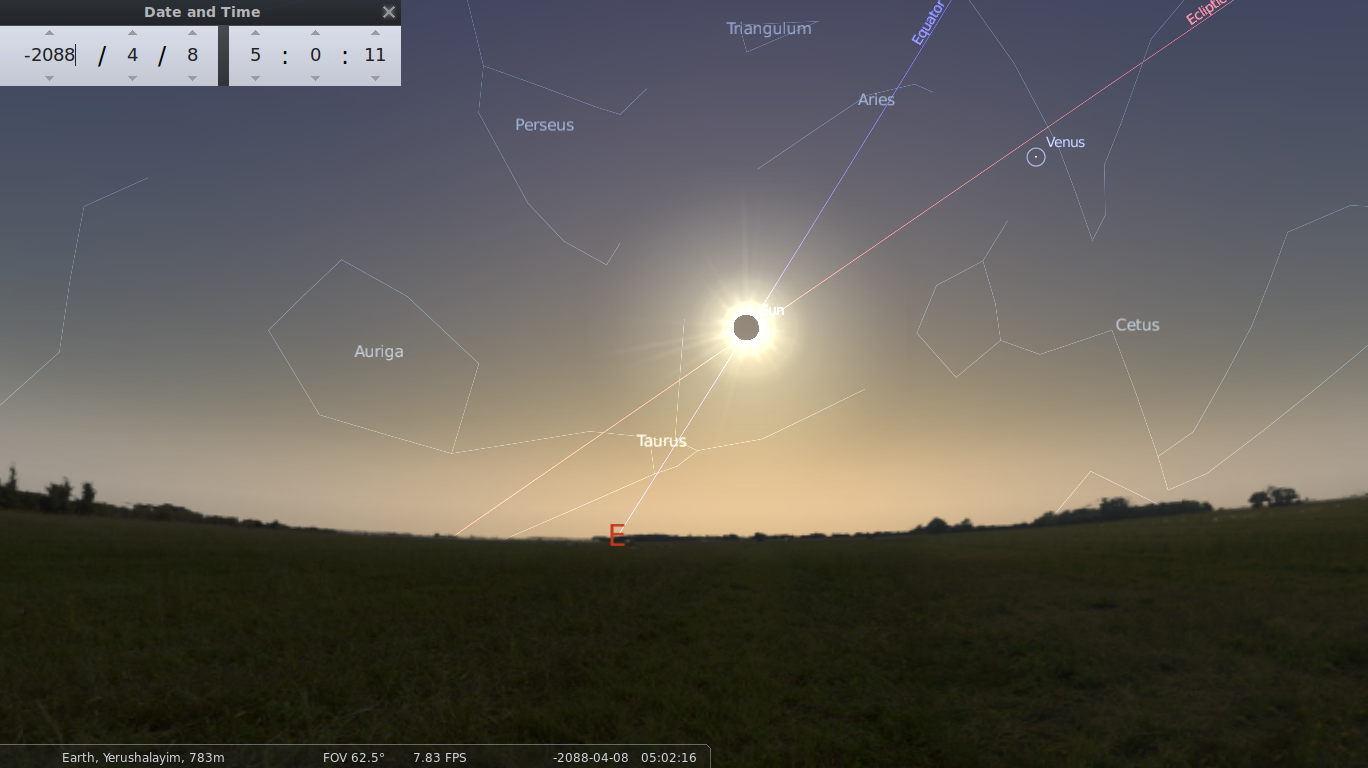}
\end{center}

\begin{quote}
-2088 4/8 

Northern equinox, total eclipse at sunrise. The Sun runs with the ecliptic (the sword), carrying along Taurus (the cattle) and the Pleiades (the daughters and children). The Moon (the hand of Laban) catches the Sun (Jacob) in the east. \end{quote} 

The omen takes place 11 years after the -2099 eclipse which simbolically begins Jacob's flight (see Gen 32:22-23), and 19 years after the -2107 eclipse which marks Jacob's entrance in the house of Laban (see Gen 31:41, ``Thus have I been twenty years in thy house``).\footnote{Observe that, in Jacob's narrative, years are explicitly reckoned as days: "And Jacob served seven years for Rachel; and they seemed unto him [but] a few days, for the love he had to her." (Gen 29:20). Similarly (as one learns seven verses later), to Laban seven years are nothing but a week: "Fulfil her week, and we will give thee this also for the service which thou shalt serve with me yet seven other years." (Gen 29:27).}  

\bigskip

\begin{verse}
Gen 31:51-53 And Laban said to Jacob, Behold this heap, and behold [this] pillar, which I have cast betwixt me and thee; This heap [be] witness, and [this] pillar [be] witness, that I will not pass over this heap to thee, and that thou shalt not pass over this heap and this pillar unto me, for harm. The God of Abraham, and the God of Nahor, the God of their father, judge betwixt us. And Jacob sware by the fear of his father Isaac.\end{verse} 

\begin{center}
 \includegraphics[width=340pt,keepaspectratio=true]{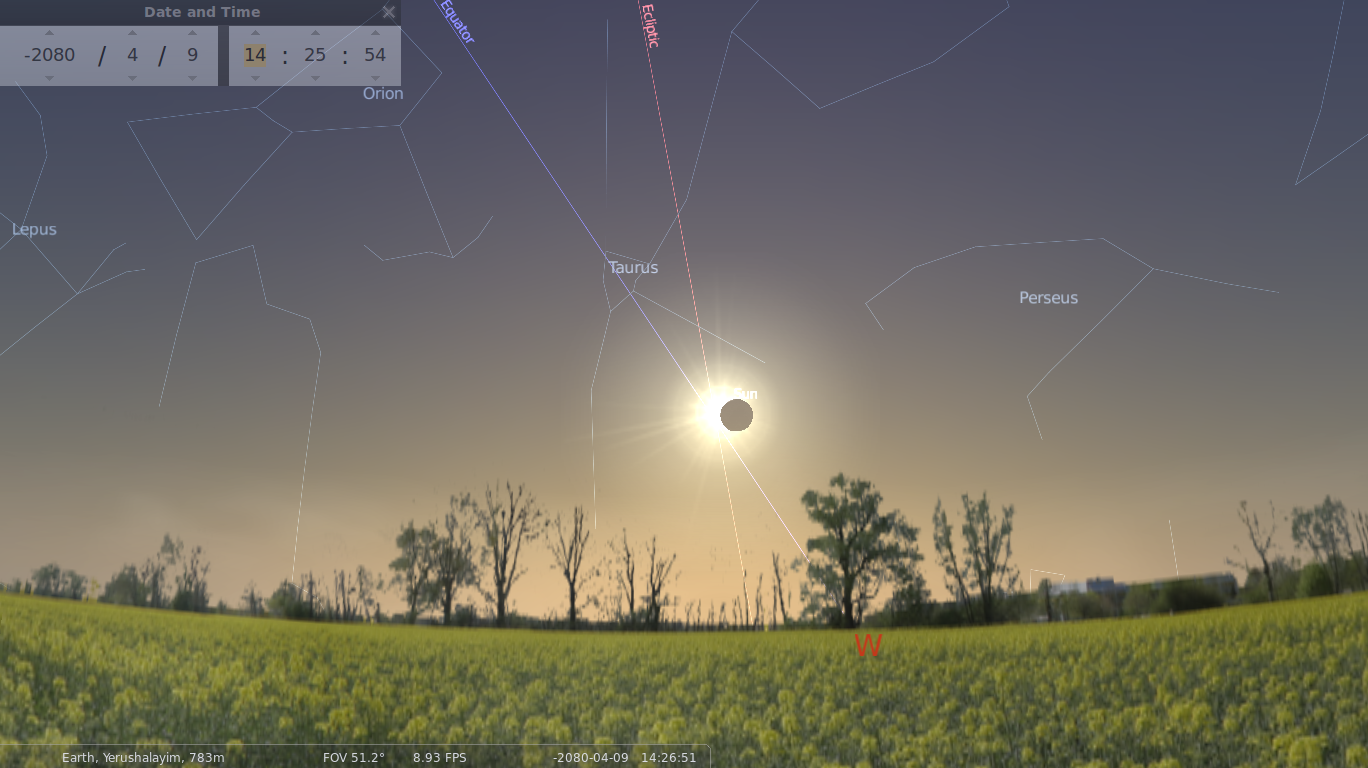}
\end{center}

\begin{quote}
-2080 4/9

Northern equinox. Partial eclipse at sunset. Covenant of Jacob with Laban in the west. The eclipse starts in daytime, so Laban swears by the God of Abraham, but prolongs into nighttime, so Jacob swears by the Fear of Isaac.                                                                                                                                                                                                                                   \end{quote} 

\bigskip

\begin{verse}
Gen 32:20-21 And say ye moreover, Behold, thy servant Jacob [is] behind us. For he said, I will appease him with the present that goeth before me, and afterward I will see his face; peradventure he will accept of me. So went the present over before him: and himself lodged that night in the company.\end{verse} 

\begin{verse}
Gen 32:24-25 And Jacob was left alone; and there wrestled a man with him until the breaking of the day. And when he saw that he prevailed not against him, he touched the hollow of his thigh [\textit{lit.}: side].\end{verse} 

\begin{verse}
Gen 32:28 And he said, Thy name shall be called no more Jacob, but Israel: for as a prince hast thou power with God and with men, and hast prevailed.\end{verse} 

\begin{verse}
Gen 32:30-31 And Jacob called the name of the place Peniel: for I have seen God face to face, and my life is preserved. And as he passed over Penuel the sun rose upon him, and he halted upon his thigh [\textit{lit.}: side].\end{verse}

\begin{quote}
-2069/4/9

Northern equinox. Close passage at nighttime. The Sun (Jacob) runs by night along the ecliptic, preceded by the constellations of the Zodiac (the presents). The Moon (the hand of God) appears poised to take over the Sun but instead it halts by its side, just before sunrise. 
\end{quote} 

\bigskip

\begin{verse}
Gen 35:9-10 And God appeared unto Jacob again, when he came out of Padanaram, and blessed him. And God said unto him, Thy name [is] Jacob: thy name shall not be called any more Jacob, but Israel shall be thy name: and he called his name Israel.\end{verse} 

\begin{quote}
-2061/4/10

Northern equinox. Annular eclipse before sunset, predicted by the Metonic cycle, when Jacob is 116 years old, just before the birth of Benjamin and the death of Rachel (Gen 35:19).\end{quote} 

Jacob is 130 years old in -2047, when he meets the Pharaoh (Gen 47:9). Jacob dies in -2030, when he is 147 years old (Gen 47:28). Observe that there are in total ten events associated with Abraham (including the one mentioned in Gen 12:1), four nighttime events associated with Isaac (one of which is shared with Abraham), and six events associated with Jacob. Not only they account for all the Paschal eclipses which took place in the timeline of the narrative (see Appendix 1): they also explain all the occurrences of the words ''appeared`` and ''covenant`` in the corresponding portions of the text.

\section{Joseph in Egypt}

\begin{verse}
Gen 41:46 And Joseph [was] thirty years old when he stood before Pharaoh king of Egypt. And Joseph went out from the presence of Pharaoh, and went throughout all the land of Egypt.\end{verse} 

Joseph is 30 years old in -2063, when he first stands in front of the Pharaoh. In that period, the Eleventh dynasty (2134-1991 BCE) ruled in Thebes. 
\bigskip

In the 22th century BCE, a severe aridification event (the 4.2 kyear event) contributed to the collapse of the Old Kingdom in Egypt and the Akkadian empire in Mesopotamia [7, 8], and was followed by a long period of famine, strife and instability, which in Egypt corresponded to the First Intermediate Period. It was during the Eleventh dynasty that the First Intermediate Period came to an end, and all of ancient Egypt was united under the Middle Kingdom [9]. 

\bigskip

\begin{verse}
Gen 41:56-57 And the famine was over all the face of the earth: And Joseph opened all the storehouses, and sold unto the Egyptians; and the famine waxed sore in the land of Egypt. And all countries came into Egypt to Joseph for to buy [corn]; because that the famine was [so] sore in all lands.\end{verse} 

\bigskip




\begin{verse}
Gen 46:33-34 And it shall come to pass, when Pharaoh shall call you, and shall say, What [is] your occupation? That ye shall say, Thy servants' trade hath been about cattle from our youth even until now, both we, [and] also our fathers: that ye may dwell in the land of Goshen; for every shepherd [is] an abomination unto the Egyptians.\end{verse} 

\begin{verse}
Gen 47:5-6 And Pharaoh spake unto Joseph, saying, Thy father and thy brethren are come unto thee:
The land of Egypt [is] before thee; in the best of the land make thy father and brethren to dwell; in the land of Goshen let them dwell.\end{verse} 

\begin{verse}
Gen 47:11 And Joseph placed his father and his brethren, and gave them a possession in the land of Egypt, in the best of the land, in the land of Rameses, as Pharaoh had commanded.\end{verse} 

\begin{verse}
Gen 47:27 And Israel dwelt in the land of Egypt, in the country of Goshen; and they had possessions therein, and grew, and multiplied exceedingly.\end{verse} 

\bigskip

Joseph dies in -1983, when he is 110 years old (Gen 50:22).

\begin{verse}
Exd 1:6 And Joseph died, and all his brethren, and all that generation.                                                                       \end{verse} 

\begin{verse}
Exd 1:8-11 Now there arose up a new king over Egypt, which knew not Joseph.
And he said unto his people, Behold, the people of the children of Israel [are] more and mightier than we:
Come on, let us deal wisely with them; lest they multiply, and it come to pass, that, when there falleth out any war, they join also unto our enemies, and fight against us, and [so] get them up out of the land.
Therefore they did set over them taskmasters to afflict them with their burdens. And they built for Pharaoh treasure [\textit{lit.}: fortified] cities, Pithom and Raamses.\end{verse} 

The enemies were probably the Hyksos, the reviled ``shepherd kings''. The Hyksos were Semitic invaders who, by the end of the Twelfth dynasty, had taken control of the delta region, extending their dominion to the whole of Lower Egypt during the Fifteenth dynasty. They were expelled during the Seventeenth dynasty, settling in Canaan but continuing to put pressure on the Egyptian borders [10]. 

Raamses, one of the two fortified cities built by the Israelites, is believed to have been founded by Paramesse while he served as vizier for Horemheb [11], the last Pharaoh of the Eighteen dynasty. Horemheb, being childless, designated as his successor Paramesse, who became the first Pharaoh of the 19th dynasty with the name Ramesses I.







\section{Moses in Midian}

\begin{verse}
Exd 2:15-17 Now when Pharaoh heard this thing, he sought to slay Moses. But Moses fled from the face of Pharaoh, and dwelt in the land of Midian: and he sat down by a well. Now the priest of Midian had seven daughters: and they came and drew [water], and filled the troughs to water their father's flock. And the shepherds came and drove them away: but Moses stood up and helped them, and watered their flock.\end{verse} 


\begin{quote}
-1355/4/4

A close passage at the Vernal equinox. As the Sun rises, the Moon appears like a dark circle in the Sun's aura, which slowly turns towards the Pleiades (the seven sisters) as the Sun proceeds in the direction of Aries (the shepherds).
\end{quote} 

The Pharaoh at the time of Moses' sojourn in Midian was Amenhotep (Amenophis) III, the ninth pharaoh of the Eighteenth dynasty (reigned from 1386 BCE to 1349 BCE, or from 1388 BCE to 1351 BCE). His son, Akhenaten, was known before the fifth year of his reign as Amenhotep IV. Akhenaten ruled for 17 years, and died in 1336 BCE or 1334 BCE. He famously abandoned traditional Egyptian polytheism in favor of a system of beliefs centered on the Aten, regarded by some as an early form of monotheism. 

\bigskip

\begin{verse}
Exd 3:1-2 Now Moses kept the flock of Jethro his father in law, the priest of Midian: and he led the flock to the backside of the desert, and came to the mountain of God, [even] to Horeb. And the angel of the LORD appeared unto him in a flame of fire out of the midst of a bush [\textit{lit.}: thorn]: and he looked, and, behold, the bush burned with fire, and the bush [was] not consumed.
\end{verse} 

\begin{quote}
-1336/4/4 

Northern equinox, close passage at sunrise. The Moon stands besides the Sun, which appears to set the eastern horizon aflame, while rising over the Gulf of Aqaba (a thorn). 
\end{quote} 

\section{Moses returns to Egypt}

\begin{verse}
Exd 4:19 And the LORD said unto Moses in Midian, Go, return into Egypt: for all the men are dead which sought thy life.
\end{verse} 

\bigskip

Tutankhamun succeeded Akhenaten at the age of eight or nine, and died while still in his teens, in 1326 or 1324 BCE. During his brief reign he was assisted by Akhenaten's own vizier, Ay, and head of the army, Horemheb. In opposition to Akhenaten's religious innovations, Tutankhamun reestablished traditional Egyptian polytheism. Tutankhamun was succeeded by Ay, who only reigned for four years. Horemreb, who succeeded Ay, was the last Pharaoh of the 18th dynasty, and reigned from 1321 BCE to 1294 BCE. or from 1319 BCE to 1292 BCE. He ``appointed judges and regional tribunes'', ``reintroduced local religious authorities'' [12].




\bigskip

\begin{verse}
Exd 4:24-26 And it came to pass by the way in the inn [\textit{lit.}: caravan circle], that the LORD met him, and sought to kill him. Then Zipporah took a sharp stone, and cut off the foreskin of her son, and cast [it] at his feet, and said, Surely a bloody husband [art] thou to me. So he let him go.\end{verse} 

\begin{center}
 \includegraphics[width=340pt,keepaspectratio=true]{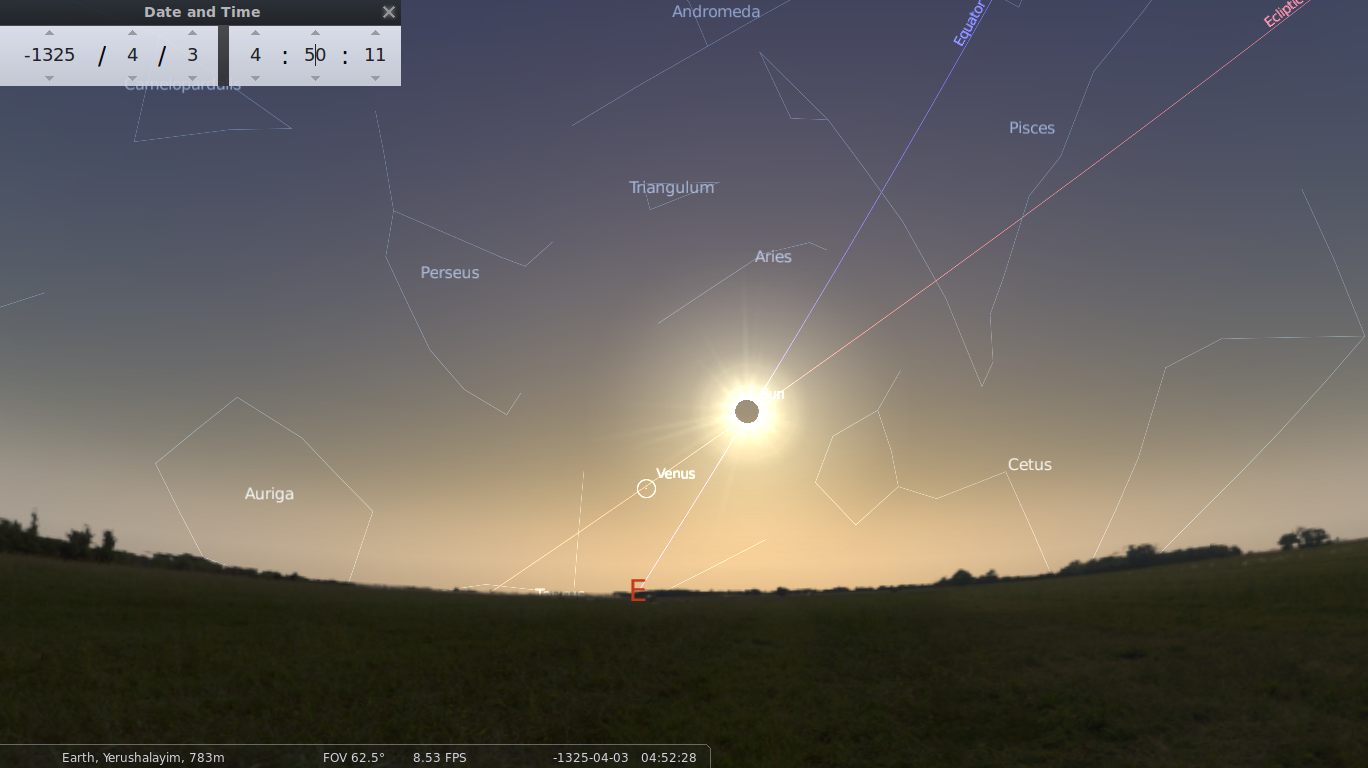}
\end{center}

\begin{quote}
-1325/4/3

At the Norhern equinox, the Sun is born in an annular eclipse. The Moon cuts out a ring from the Sun's disc, which becomes the only part visible from the ground. \end{quote} 

\bigskip

Iah (also spelled Yah, Jah, or Aah) was an Egyptian moon god, and also the Egyptian word for Moon [13].

\section{Passover of 1318 BCE: Exodus from Egypt}

\begin{verse}
Exd 12:2 This month [shall be] unto you the beginning of months: it [shall be] the first month of the year to you.                                                                                                                  \end{verse} 

\begin{verse}
Exd 12:14 And this day shall be unto you for a memorial; and ye shall keep it a feast to the LORD throughout your generations; ye shall keep it a feast by an ordinance for ever.
\end{verse} 

\begin{verse}
Exd 12:21 Then Moses called for all the elders of Israel, and said unto them, Draw out and take you a lamb according to your families, and kill the passover.
\end{verse}

\begin{verse}
 Exd 12:6 And ye shall keep it up until the fourteenth day of the same month: and the whole assembly of the congregation of Israel shall kill it in the evening.
\end{verse}

\begin{verse}
Exd 12:29 And it came to pass, that at midnight the LORD smote all the firstborn in the land of Egypt.
\end{verse} 


\begin{verse}
-1317/4/4

Nighttime, partial eclipse at the Northern equinox. The Sun (the firstborn) in Aries (a lamb) is overtaken by the Moon (the hand of God) around midnight.
\end{verse} 

The Exodus began with the symbolic nighttime slaughter of the Sun in Aries: in 1318 BCE the Israelites, led by Moses, fled from Egypt. 
Observe that our dating of the Exodus falls within seven years of the traditional date of 1312-1311 BCE from the Seder Olam Rabbah. The seven-year gap can perhaps be explained by the fact that the traditional date was computed from the construction of the First Temple (according to the Talmud, 410 years prior to its destruction in 422-421 BCE), and that the Temple took seven years to complete (1 Kings 6:38). If the traditional date of 832-831 BCE for the construction of the Temple is interpreted as the date of its completion, then the work started in 839-838 BCE. Since Solomon began building it 480 years after the Exodus (1 Kings 6:1), then the latter can be dated to 1319-1318 BCE, providing an external validation for the date obtained from the internal chronology of the Pentateuch.

\bigskip

The Egyptian bondage symbolically starts with the death of Joseph in -1983 (Exd. 1:8), and lasts until -1317, the year of the Exodus. Hence, in the internal chronology, the Egyptian bondage lasts exactly 666 years.

\bigskip

\begin{verse}
Exd 13:4 This day came ye out in the month Abib [\textit{lit.}: Ripening].
\end{verse}

\begin{verse}
 Exd 13:10 Thou shalt therefore keep this ordinance in his season from year to year.
\end{verse}

The Egyptian calendar was based on a 365-day year, divided into three seasons of four months each: Inundation (Akhet), Growth (Proyet), and Harvest (Shemu), followed by an additional intercalary month of just 5 days. Each regular month had 30 days, and was simply denoted by its position in the season, from first to fourth. The Harvest season began exactly 240 days after the heliacal rising of Sirius, as observed from a conventional location, which every year marked the beginning of the Inundation season. In the 14th century BCE the event would have been observed in the second half of July. If the Exodus began in the first month of the Harvest season, on the 15th day of the month, this implies a date of of July 19, 1319 BCE for the heliacal rising of Sirius, as recorded from the appointed location.

\section{The Isrealites in the wilderness}

\begin{verse}
Exd 14:19-21 And the angel of God, which went before the camp of Israel, removed and went behind them; and the pillar of the cloud went from before their face, and stood behind them: And it came between the camp of the Egyptians and the camp of Israel; and it was a cloud and darkness to them, but it gave light 
by night to these: so that the one came not near the other all the night. And Moses stretched out his hand over the sea; and the LORD caused the sea to go back by a strong east wind all that night, and made the sea dry land, and the waters were divided.
\end{verse} 

\begin{quote}
-1317/4/18 

Total lunar eclipse, shortly after the full Moon has risen in the East. The Moon is obscured while passing over the camp of the Israelites, but shines again while passing over the Egyptian camp behind. The omen takes place exactly two weeks after the beginning of the Exodus. \end{quote}  

\begin{verse}
Exd 16:1 And they took their journey from Elim, and all the congregation of the children of Israel came unto the wilderness of Sin, which [is] between Elim and Sinai, on the fifteenth day of the second month after their departing out of the land of Egypt. 
\end{verse} 

\begin{verse}
Exd 16:10 And it came to pass, as Aaron spake unto the whole congregation of the children of Israel, that they looked toward the wilderness, and, behold, the glory of the LORD appeared in the cloud.
\end{verse} 

\begin{quote}
-1317/5/4

Partial eclipse around noon, exactly 30 days after the nighttime eclipse which marked the beginning of the Exodus. 
\end{quote} 

On the next day, -1317/5/5, manna is first collected by the Israelites:

\begin{verse}
Exd 16:13-15 And it came to pass, that at even the quails came up, and covered the camp: and in the morning the dew lay round about the host. And when the dew that lay was gone up, behold, upon the face of the wilderness there lay a small round thing, as small as the hoar frost on the ground. And when the children of Israel saw it, they said one to another, It is manna: for they wist not what it was. And Moses said unto them, This is the bread which the LORD hath given you to eat.
\end{verse}

\begin{verse}
Exd 16:22-23 And it came to pass, that on the sixth day they gathered twice as much bread, two omers for one man: and all the rulers of the congregation came and told Moses. And he said unto them, This is that which the LORD hath said, To morrow is the rest of the holy sabbath unto the LORD.
\end{verse} 

Manna was thus collected for six days, while the seventh morning, -1317/5/11, was a shabbat. As it turns out, the Julian date of May 11, 1318 BCE corresponds to a Saturday. This suggests that, at least since the time of redaction, shabbat has been consistently celebrated on the same day of the week.

\bigskip

The sojourn of the children of Israel in the desert lasts forty years (Exd 16:35), until -1277, when Moses dies, being 120 (Deu 34:7). Hence, Moses was born in -1397, and was 42 in -1355, when he met his wife. He was 80 years old in -1317, when he led the Israelites out of Egypt. 

\bigskip

\begin{verse}
Exd 24:10 And they saw the God of Israel: and [there was] under his feet as it were a paved work of a sapphire [\textit{lit}: sapphire-colored] stone, and as it were the body of heaven in [his] clearness.
\end{verse} 

\begin{quote}
-1306/4/2

Northern equinox, partial eclipse just before sunset. The Sun is partially covered by the Moon while setting over the Gulf of Suez. The exact location could be Wadi Maghareh, a site in southwestern Sinai known in ancient Egypt as ``the Terraces of Turquoise``. The first tables of the Law are delivered shorty after this date, when Moses is 91 years old.
\end{quote} 

\bigskip

\begin{verse}
Exd 33:18-23 And he said, I beseech thee, shew me thy glory.
And he said, I will make all my goodness pass before thee, and I will proclaim the name of the LORD before thee; and will be gracious to whom I will be gracious, and will shew mercy on whom I will shew mercy.
And he said, Thou canst not see my face: for there shall no man see me, and live.
And the LORD said, Behold, [there is] a place by me, and thou shalt stand upon a rock:
And it shall come to pass, while my glory passeth by, that I will put thee in a clift of the rock, and will cover thee with my hand while I pass by: And I will take away mine hand, and thou shalt see my back parts: but my face shall not be seen.
\end{verse} 

\begin{center}
 \includegraphics[width=340pt,keepaspectratio=true]{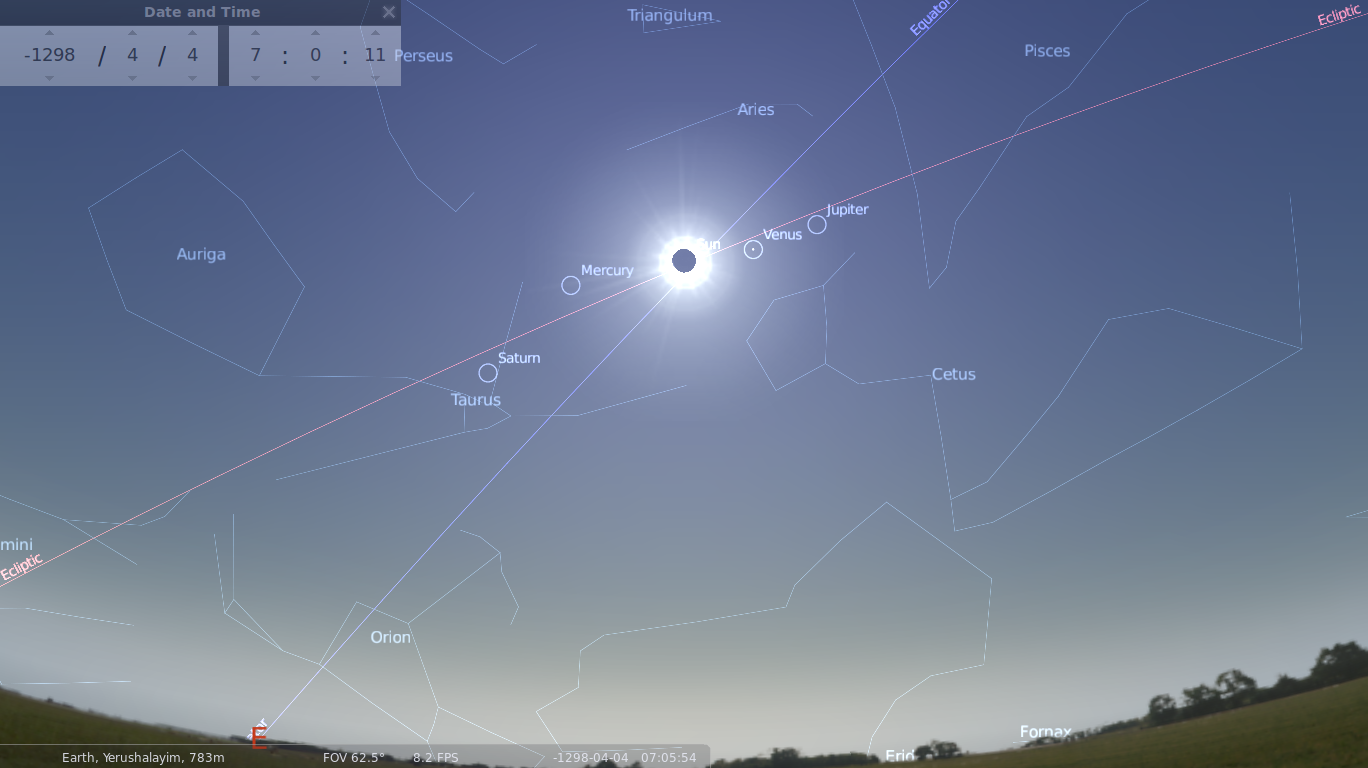}
\end{center}

\begin{quote}
-1298/4/4

Annular eclipse at the Northern equinox. The Sun rises while covered by the Moon, but sets uncovered. Venus and Jupiter precede it in its path, Mercury and Saturn follow closely. 
\end{quote} 

In -1298 Moses is 99 years old. The cleft in the rock could be Gidi Pass, a strategically important pass in the mountain region of Sinai, which extends for about 29 km along the east-west axis, or Mitla Pass, which runs in the same direction just south of the first, and is about 32 km long. Together, the two passes control all vehicle access from the desert to the mountain region of Sinai. 

\bigskip

\begin{verse}
Deu 31:10-11 And Moses commanded them, saying, At the end of [every] seven years, in the solemnity of the year of release, in the feast of tabernacles, when all Israel is come to appear before the LORD thy God in the place which he shall choose, thou shalt read this law before all Israel in their hearing.
\end{verse} 

The new tables of the Law are delivered to the people just after the Northern equinox of 1299 BCE, slightly less than eight years after the original tables were prepared.

\bigskip

\begin{verse}
Num 20:6 And Moses and Aaron went from the presence of the assembly unto the door of the tabernacle of the congregation, and they fell upon [\textit{lit.}: lowered] their faces: and the glory of the LORD appeared unto them.
\end{verse} 

\begin{quote}
-1287/4/2 

Nighttime close passage at the Northern equinox. The Moon appears to closely align with the Sun, while the latter is still below the eastern horizon. At the time of the omen Moses was exactly 110 years old.
\end{quote} 

\bigskip

\begin{verse}
Num 22:21-22 And Balaam rose up in the morning, and saddled his ass, and went with the princes of Moab. And God's anger was kindled because he went: and the angel of the LORD stood in the way for an adversary against him. Now he was riding upon his ass, and his two servants [were] with him.
\end{verse} 

\begin{verse}
Num 22:31 Then the LORD opened the eyes of Balaam, and he saw the angel of the LORD standing in the way, and his sword drawn in his hand: and he bowed down his head, and fell flat on his face.
\end{verse} 

\begin{quote}
-1279/4/3 

Annular eclipse at the Northern equinox, just after sunset. The Sun (Balaam) proceeds with the signs of the Zodiac (the princes of Moab). The Moon (the angel of God) holds on to the ecliptic (the sword), blocking the way.
\end{quote} 


In -1279 Moses is 118 years old. Observe that there are eight Paschal events (including close passages) in Moses' timeline, and that both Abraham and Moses witnessed Paschal events when they were exactly 80, 91, 99, 110, and 118 years old. 

\section{And it is now 430 years...}

\begin{verse}
Exd 12:40-41 Now the sojourning of the children of Israel, who dwelt in Egypt, [was] four hundred and thirty years. And it came to pass at the end of the four hundred and thirty years, even the selfsame day it came to pass, that all the hosts of the LORD went out from the land of Egypt.                                                                                                                                                                                                                                                                                               \end{verse} 

\begin{verse}
Exd 16:36 Now an omer [is] the tenth [part] of an ephah.\end{verse} 

\begin{quote}
-888/3/31

Total eclipse at the Northern equinox. The Sun in Aries is covered by the Moon, but eventually breaks free from its shadow.                                                                                                                                \end{quote} 

The omen was observed at the Northern equinox of 889 BCE, 429 years after the Northern equinox of 4/4 1318 BCE, which symbolically began the Exodus. Hence, this portion of the text was probably composed shortly after that date. 



\section{Back to Eden}

The genealogies of the pre-flood patriarchs widely differ across manuscripts. Three main traditions are extant: the Masoretic, the Septuagint and the Samaritan, and they all differ in their pre-flood genealogies. In all cases, setting the date of the Flood at -5500, Noah starts to reign in -6100; yet, all other dates are different in the three texts. Strikingly, while the genealogical time spans in the Masoretic and Septuagint text give rise to chronologies with no apparent connection to astronomical events, using the genealogical time spans in the Samaritan Pentateuch one finds numerous eclipses in the proximity of the Northern equinox marking the birth or beginning of the reign of the pre-flood patriarchs, in addition to several close passages of the Moon. The table below shows the reconstructed Samaritan chronology, together with the associated solar eclipses and close passages. 

\begin{center}
\begin{tabular}{llll}
Adam reigns & -13237 & - & -\\
Seth born & -13107 & -13107 & partial, 7/4, 22:45, -3\\
Seth reigns & -12307 & -12306 & total, 6/28, 8:45, -3\\
Enos born & -12202 & -12203 & annular, 6/29, 12:44, -1\\
Enos reigns & -11395 & -11394 & total, 6/25, 2:43, +1\\
Cainan born & -11305 & - & -\\
Cainan reigns & -10490 & - & -\\
Mahalaleel born & -10420 & - & -\\
Mahalaleel reigns & -9580 & -9580 & close passage, 6/8, 21:42, -1\\
Jared born & -9515 & -9515 & annular, 6/10, 7:40 +2\\
Jared reigns & -8685 & -8684 & close passage, 6/2, 12:40, +1\\
Enoch born & -8623 & - & -\\
Enoch reigns & -7838 & - & -\\
Methuselah born & -7773 & -7774 & annular, 5/23, 3:11, -3\\
Methuselah reigns & -7473 & - & -\\
Lamech born & -7406 & - & -\\
Lamech reigns & -6753 & - & -\\
Noah born & -6700 & - & -\\
Noah reigns & -6100 & -6099 & annular, 5/13, 3:09, +2
\end{tabular}
\end{center}

Observe that eclipses either occur within one year before birth (i.e., a 0 or -1 year difference), or within the first year of reign (i.e., a 0 or +1 year difference). 

Let us now proceed to test the hypothesis that the dates in the reconstructed Samaritan chronology were set independently of eclipse events. Allowing for -1 year (resp., +1 year) from the date of birth (resp., reign) of each patriarch, and allowing for +/- 3 days for each observation (i.e., selecting all eclipses which took place within 72 hours of the passage of the Sun at the Northern equinox), one finds 7 events for 19 data points. Assuming that the frequency of solar eclipses over sufficiently long periods of time is independent of the calendar date, and using the average bound of 2.4 events per year, gives a probability of solar eclipse in the proximity of the Northern equinox of 2x(2.4)x6/(365.25), which amounts to about 0.079. We can now apply a binomial probability model to estimate the chance of 7 or more successes in 19 independent trials, if the probability of each success is given by 0.079. We find that, in the case of the Samaritan chronology, the probability that the result is due to pure chance, in case the dates were set independently of eclipse events, is about 4 in 10,000. By contrast, in the case of the Masoretic and Septuagint chronologies there are only one and three eclipses, respectively, in the proximity of the dates of birth and death of the pre-flood patriarchs (in addition to a few close passages), and hence the null hypothesis cannot be rejected at any interesting level of significance.


Also observe that not all dates are set in accordance to astronomical events. This is not surprising, as some of the genealogical numbers evidently carry different types of information. Notice, for instance, the decreasing lifespans, and decreasing time before birth of heir, until Enoch, who is the seventh patriarch, and reigns for exactly 365 years (recall how, in various points in the text, years are identified with days). The sum of reignal years of the six before him is 5400, or exactly 6 x 900, a number which is divisible for 1,2,3,4,5,6. The product of those first six numbers is 720, like the reignal years of Methuselah.

%
%
%
%
%

\bigskip

\begin{verse}
Gen 3:15 And I will put enmity between thee and the woman, and between thy seed and her seed; it shall bruise thy head, and thou shalt bruise his heel.
\end{verse} 

\begin{center}
 \includegraphics[width=340pt,keepaspectratio=true]{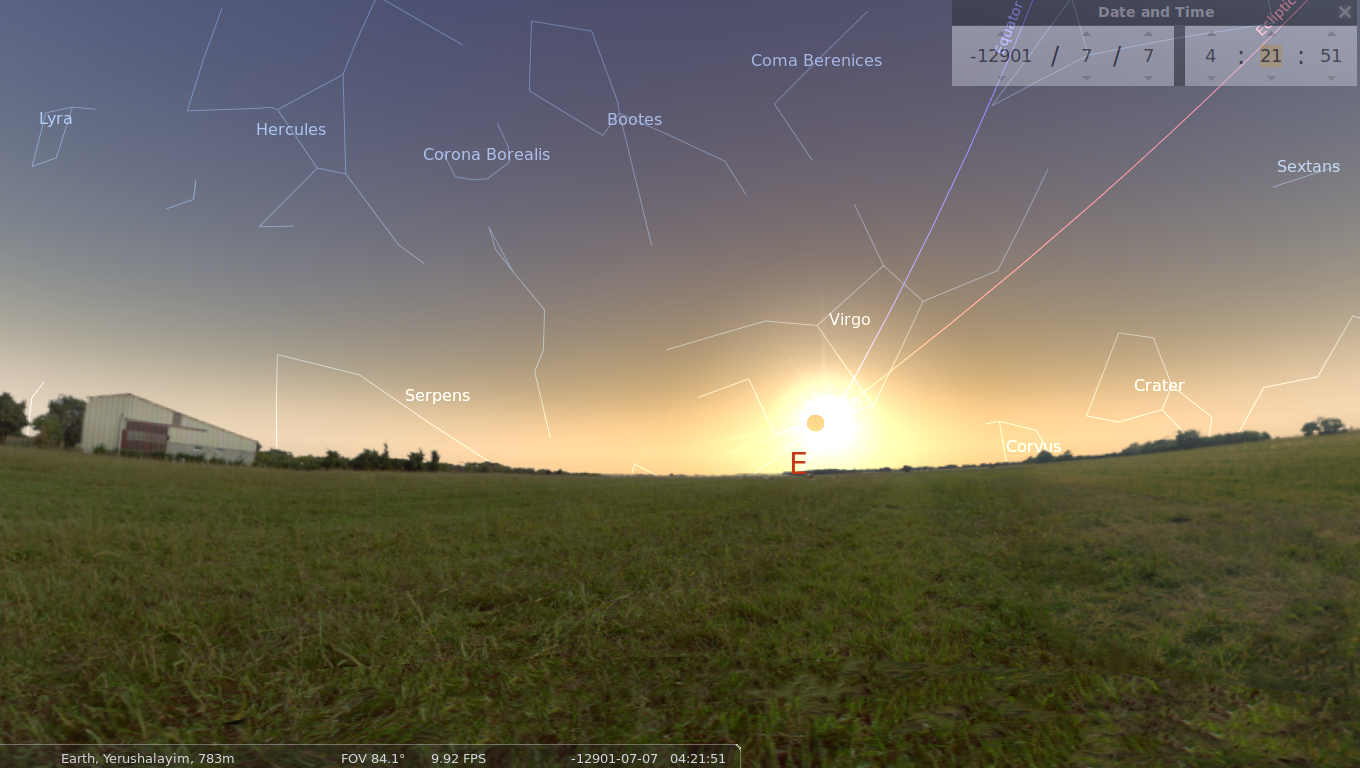}
\end{center}

\begin{quote}
-12901/7/7

Northern equinox, partial eclipse at sunrise. The Sun, standing on the leg of Virgo (the Woman), is obscured by the Moon, while Virgo's other leg points towards Serpens' head.
\end{quote} 

\bigskip

When Adam is driven out from the valley of Eden, he starts to till the land (Gen 3:23 ``Therefore the LORD God sent him forth from the garden of Eden, to till the ground from whence he was taken.''). The agricultural revolution begins in the Near East around 12000 BCE, while the world was experiencing a dramatic climate event: the rapid rise of the level of the oceans, which at the time was about 100 meters lower than today, by more than 20 meters within just two centuries (Meltwater Pulse 1A) [14], and the gradual flooding of the most fertile plains and coastlines, starting around 12700 BCE. The most prominent loss in the Near East was that of the Persian Gulf, which at the time was a vast alluvial plain, fertilized by the combined waters of the Tigris and Euphrates rivers. As the level of the oceans increased the Persian Gulf became a marshland, and was eventually submerged. 

Roughly contemporary with Meltwater Pulse 1A, and perhaps triggered by it, came a sudden cooling of the global temperature (Older Dryas / Antartic Cold Reversal), whose onset is estimated at around 12500 BCE.

\bigskip




\begin{verse}
Gen 3:21 Unto Adam also and to his wife did the LORD God make coats of skins, and clothed them.                                                                                               \end{verse} 

\begin{verse}
Gen. 3:24 So he drove out the man; and he placed at the east of the garden of Eden Cherubims, and a flaming sword which turned every way, to keep the way of the tree of life.
\end{verse} 

The root C-R-B(P) is attested in many languages to represent a CaRaPax, or any creature endowed with a carapace: a CRaB, a sCaRaB, a sCoRPion, a CRay(B)fish, even metaphorically (an armored warrior).

\begin{center}
 \includegraphics[width=340pt,keepaspectratio=true]{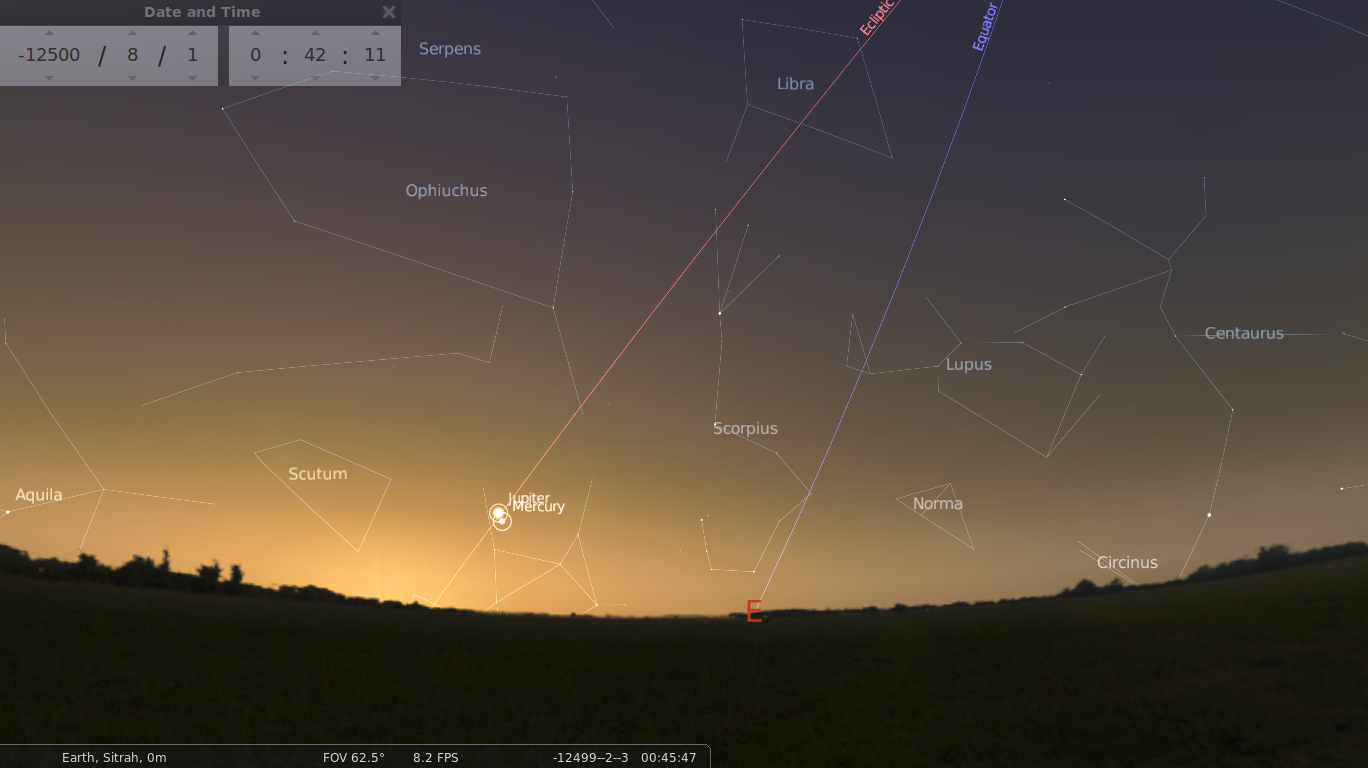}
\end{center}

\begin{quote}
-12700 to -12500, Norhern solstice, sunrise.

In this time span, at the Northern (Summer) solstice one would see Scorpio rising in the east just before sunrise, holding the revolving, ``flaming sword'' of the ecliptic.                                                                                                                                                                            \end{quote} 


\begin{verse}
 Isa 26:4 Trust ye in the LORD for ever: for in the LORD JEHOVAH [is an] everlasting strength [\textit{lit.}: rock].
\end{verse}

\begin{verse}
Deu 32:4 [He is] the Rock, his work [is] perfect.\end{verse}

If measured from the time of the Curse, the current year 2011 in the Gregorian calendar corresponds to year 14912 in the internal chronology of the Pentateuch. It is a calendar of the neolithic Near East, attuned to those events and places by both human time-keeping, based on counting days and years, and divine time-keeping, based on the unique positions and cycles of the celestial bodies.

\section{References}

[1] Elrington, Charles Richard (1847), The Whole Works of the Most Rev. James Ussher, D.D., XIII, Dublin: Hodges and Smith.

[2] A. Aaboe (1974), "Scientific Astronomy in Antiquity", Philosophical Transactions of the Royal Society 276 (1257): 21-42.

[3] Ryan, W.B.F., Pitman III, W.C., et al. (1997) ``An abrupt drowning of the Black Sea shelf''. Marine Geology, 138: 119-126.

[4] Ryan, W.B.F. and Pitman III, W.C. (1999) Noah's Flood: The New Scientific Discoveries About the Event That Changed History. Simon \& Schuster.

[5] Botsforth, George W., ed. "The Reign of Sargon". A Source-Book of Ancient History. New York: Macmillan, 1912.

[6] Grayson, Albert Kirk. Assyrian and Babylonian Chronicles. J. J. Augustin, 1975; Eisenbrauns, 2000.

[7] Bar-Matthews, Miryam; Ayalon, Avner; Kaufman, Aaron (1997). "Late Quaternary Paleoclimate in the Eastern Mediterranean Region from Stable Isotope Analysis of Speleothems at Soreq Cave, Israel". Quaternary Research 47 (2): 155-168.

[8] deMenocal, Peter B. (2001), "Cultural Responses to Climate Change During the Late Holocene", Science 292 (5517): 667-673.

[9] Juergen von Beckerath, The Date of the End of the Old Kingdom of Egypt, JNES 21 (1962), p.146

[10] Grimal, Nicolas. A History of Ancient Egypt, p.193. Librairie Artheme Fayard, 1988.

[11]  K.A. Kitchen, On the Reliability of the Old Testament. William B. Eerdmans Co., 2003. p.255

[12] Erik Hornung, Rolf Krauss and David Warburton (editors), Ancient Egyptian Chronology (Handbook of Oriental Studies), Brill: 2006, p.493

[13] Allen, James P. (2000). Middle Egyptian: An Introduction to the Language and Culture of Hieroglyphs. Cambridge University Press. p. 436

[14] Stanford, Jennifer D.; Rohling, Eelco J.; Hunter, Sally E.; Roberts, Andrew P.; Rasmussen, Sune O.; Bard, Edouard; McManus, Jerry; Fairbanks, Richard G. (9 December 2006). "Timing of meltwater pulse 1a and climate responses to meltwater injections". Paleoceanography (American Geophysical Union) 21: 4103.

[15] Hachlili, Rachel. Ancient Jewish Art and Archaeology in the Land of Israel. Brill: 1987, p.288 

\section{Appendix 1: All solar eclipses and close passages at the Northern equinox from -2862 to -2061}

In the table below the approximate time of day is for a reference point (observer) located in the Near East, but the classification (annular, total, partial, or close passage) is global. 
Close passages are those where the Sun and the Moon disks do not overlap, but appear within one solar diameter from each other. All eclipses and close passages in the proximity of the Northern equinox (+/-3 days, i.e., +/- 72 hours from the time of the equinox) in the eight-century period between -2862 and -2061 are represented, in addition to a few other notable events in the timeline of the narratives. In particular, the event on -2422/4/11 is not a close passage, but is still notable due to the fact that, at the Northern equinox, the new Moon appears to touch the Pleiades. Finally, for each event we also report whether it was linked to a preceding event through the 19-year Metonic cycle.

\begin{center}
\begin{tabular}{lll}
-2862 & 4/15, +1 & Close passage just after sunrise\\
-2851 & 4/13, -1 & Total eclipse, daytime\\
-2843 & 4/14, +1 & Partial eclipse, nighttime\\
-2832 & 4/13, -1 & Close passage, sunset\\
-2824 & 4/14, +2 & Total eclipse, nighttime\\
-2805 & 4/16, +2 & Total eclipse, nighttime\\
-2797 & 4/17, +3 & Close passage, sunrise\\
-2786 & 4/15, +2 & Annular eclipse, nighttime\\
-2767 & 4/15, +2 & Close passage, daytime\\ 
-2721 & 4/16, +3 & Annular eclipse, daytime\\
-2702 & 4/16, +3 & Partial eclipse, daytime\\
-2620 & 4/9, -3 & Close passage, nighttime\\
-2601 & 4/10, -2 & Partial eclipse, nighttime\\
-2582 & 4/10, -2 & Annular eclipse, daytime\\
-2563 & 4/9, -3 & Annular eclipse, nighttime\\
-2544 & 4/9, -3 & Annular eclipse, nighttime\\
-2536 & 4/10, -1 & Close passage, nighttime\\
-2525 & 4/9, -3 & Partial eclipse, sunset\\
-2517 & 4/11, -1 & Annular eclipse, daytime\\
-2506 & 4/9, -3 & Close passage, daytime\\
-2498 & 4/11, -1 & Annular eclipse, nighttime\\
-2471 & 4/12, +1 & Partial eclipse at sunrise, Abraham\\
-2452 & 4/12, +2 & Total eclipse at sunrise, Metonic, Abraham\\
-2441 & 4/11, +1 & Close passage at nighttime, Abraham\\
-2433 & 4/13, +2 & Annular eclipse at sunrise, Metonic, Abraham\\
-2422 & 4/11, +1 & New Moon touches Pleiades, Metonic, Abraham\\
-2414 & 4/12, +2 & Annular, nighttime, Metonic, Abraham and Isaac\\
-2406 & 4/14, +4 & Close passage, just before sunset, Abraham\\
-2395 & 4/12, +2 & Partial, Metonic, Abraham\\
-2387 & 4/14, +4 & Annular, sunrise, Metonic, Abraham\\
-2368 & 4/13, +4 & Annular, sunset, Metonic, Abraham\\
-2349 & 4/13, +4 & Annular, nighttime, Metonic, Isaac\\
-2330 & 4/13, +4 & Partial, daytime, Metonic\\
-2311 & 4/13, +4 & Close passage, daytime, Metonic\\
-2303 & 4/14, +5 & Total, nighttime, Isaac\\
-2284 & 4/14, +6 & Total, after sunset, Metonic, Isaac\\
-2229 & 4/7, -2 & Close passage, nighttime\\
-2218 & 4/5, -4 & Partial, nighttime\\
-2210 & 4/7, -2 & Partial, sunset, Metonic\\
-2191 & 4/6, -3 & Annular, nighttime, Metonic\\
-2172 & 4/6, -3 & Annular, nighttime, Metonic\\
-2153 & 4/6, -3 & Total, nighttime, Metonic\\
-2145 & 4/8, -1 & Partial, daytime\\
-2126 & 4/8, -1 & Total, sunrise, Metonic\\
-2107 & 4/8, +1 & Total, sunrise, Metonic, Jacob\\
-2099 & 4/9, +2 & Close passage, sunset, Jacob\\
-2088 & 4/8, +1 & Total, sunrise, Metonic, Jacob\\
-2080 & 4/9, +2 & Partial, sunset, Metonic, Jacob\\
-2069 & 4/9, +1 & Close passage, before sunrise, Metonic, Jacob\\
-2061 & 4/10, +2 & Annular, sunset, Metonic, Jacob
\end{tabular}
\end{center}

\end{document}